\documentclass[twocolumn,showpacs,aip]{revtex4}
\usepackage{amssymb}
\usepackage{amsmath}
\usepackage{graphicx}
\usepackage{bm}
\usepackage{epstopdf}
\begin{document}

\title{Far-field constant-gradient laser accelerator of electrons in an ion channel}

\author{ Vladimir Khudik,  Xi Zhang, and Gennady Shvets}
\affiliation{{The University of Texas at Austin, Austin, Texas 78712, USA} }
\begin{abstract}
We predict that electrons in an ion channel can gain ultra-relativistic energies by simultaneously interacting with a laser pulse and, counter-intuitively, with a decelerating electric field. The crucial role of the decelerating field is to maintain high-amplitude betatron oscillations, thereby enabling constant rate energy flow to the electrons via the direct laser acceleration mechanism. Multiple harmonics of the betatron motion can be employed. Injecting electrons into a decelerating phase of a laser wakefield accelerator is one practical implementation of the scheme.
\end{abstract}

\maketitle

\section{Introduction}
Far-field accelerators occupy an important niche in the field of advanced electron accelerators. Such accelerators utilize a transverse electromagnetic wave that resonantly interacts with electrons undergoing transverse motion.
Examples of such transverse motion include electron undulation in
the transverse magnetic field, electron gyration in the
longitudinal magnetic field, or the betatron motion in the
confining focusing channel. The corresponding accelerator schemes
are, respectively, referred to as inverse free-electron laser
FEL~\cite{palmer_jap72,babzien_prl98,steenbergen_prl96}, cyclotron resonance laser (CRL)~\cite{kolom_jetp63,sprangle_ieee83,friedland_jap91}, and
ion-channel laser (ICL)~\cite{whittum_prl90,pukhov_dla,gahn_prl}.

The main advantage of the far-field inverse laser schemes is their simplicity: no electromagnetic structure is required because the
acceleration is accomplished by the $W_{\parallel} \propto
\vec{v}_{\perp} \times \vec{B}_{\perp}^{(L)}$ force directly
exerted on the electron beam by the transverse component of the
laser's electromagnetic fields $(\vec{E}^{(L)},\vec{B}^{(L)})$.
The main drawback is that, in general, the acceleration gradient
$W_{\parallel}$ tends to decrease as the relativistic electron
energy $\gamma mc^2$ increases. For example, $W_{\parallel}
\propto 1/\gamma$ for inverse FELs and $W_{\parallel} \propto
1/\sqrt{\gamma}$~\cite{sprangle_ieee83} for inverse CRLs, thereby
reducing the usefulness of these schemes to modest $\gamma$'s. Such reduction is caused by rapid decrease of $|\vec{v}_{\perp}|$  and cannot be cured by simply maintaining the wave-particle synchronism~\cite{sprangle_ieee83,friedland_jap91}. Therefore, it can only be overcome if a suitable mechanism of steadily increasing the magnitude of the transverse electron momentum $\vec{p}_{\perp} = \gamma m \vec{v}_{\perp}$, as well as its phase with respect to the laser field $\vec{E}^{(L)}$, can be found.

In this paper we propose that both requirements can be
satisfied for an inverse ICL if a small constant {\it
decelerating} longitudinal electric field is applied to the
electrons undergoing betatron motion inside an ion channel. The
channel can be produced by the space charge of the electron beam
itself~\cite{whittum_prl90,rosenzweig_PRA91}, or by the
ponderomotive pressure of an ultra-intense laser pulse that
creates a plasma "bubble"~\cite{pukhov_bubble}. Practical
realizations of the inverse ICL, where electrons are externally
injected into the leading (decelerating) part of the plasma
bubble, are suggested and analyzed using particle-in-cell (PIC)
simulations. The relativistic electron dynamics is analytically reduced to that of a nonlinear pendulum.

Earlier work on direct laser acceleration
(DLA)~\cite{pukhov_dla,gahn_prl,cipiccia,suk_pop,shaw_ppcf,zhang_prl}
demonstrated that a relatively small energy from the laser,
$\Delta A_L$, can be added to that gained from the {\it
accelerating} longitudinal electric field, $\Delta A_W >0$, under
special electron injection conditions and laser pulses formats:
large initial transverse energy of the injected electrons, and the elongated, asymmetric, or multi-peaked laser
pulse~\cite{suk_pop,shaw_ppcf,zhang_prl}. The unique feature of
the acceleration mechanism proposed in this paper is that the
small longitudinal electric field actually {\it reduces} the
energy of the beam by $\Delta A_W <0$, but in so doing modifies
the transverse electron dynamics to ensure that the direct energy
gain is twice as high: $\Delta A_L \approx 2|\Delta A_W|$.
Moreover, the above stringent conditions on the laser and electron beam need not apply.

The rest of the paper is organized as follows. In Sec. II we introduce the minimal model of the electron acceleration  in the ion channel under combined action of the oscillating laser electromagnetic wave and the stationary longitudinal decelerating electric field. We qualitatively analyze and numerically solve the relativistic equations of electron motion. In Sec.~III we develop analytical theory drawing analogy with motion of the pendulum. 
Section~IV presents PIC simulations confirming the developed theory. Conclusion is given in Sec.~V.

\section{Model}\label{sec:eq_motion}
We start by introducing a simple model of the electron motion in
the combined focusing field $\vec{F}_{\perp}=-m\omega_p^2
\vec{r}_{\perp}/2$ of a cylindrical ion channel (where $\omega_p = \sqrt{4\pi e^2 n/m}$ is the plasma frequency, $n$ is the plasma
density, $m$ is the electron mass, and $\vec{r}_{\perp} = y \vec{e}_y + z\vec{e}_z$ is the transverse electron's position), the electromagnetic field of a linearly ($y$-)polarized laser beam propagating with the phase velocity $v_{\rm ph}$ in the
$x$-direction, and a constant decelerating electric field
$\vec{E}_{dec} = \vec{e}_x E_{\parallel}$.  The planar laser
fields are assumed in the form of  $E_y^{(L)} = E_0
\cos{\varphi}$ and $B_z^{(L)} = c E_y^{(L)}/v_{\rm ph}$, where
$\varphi=\omega_L (x/v_{\rm ph} - t)$ is the laser phase. The
relativistic equations of motion are given by
\begin{eqnarray}
&&\frac{d}{dt} \left( \gamma \frac{dx}{dt} \right) = -\frac{e}{m} \left( E_{\parallel} + E_0
\frac{v_y \cos{\varphi}}{v_{\rm ph}} \right), \label{eq:EqM_N1} \\
&&\frac{d}{dt} \left(\gamma \frac{dy}{dt} \right) = -\frac{\omega_p^2 y}{2}
+ \frac{eE_0}{m} \left( \frac{v_x}{v_{\rm ph}} - 1 \right) \cos{\varphi}, \label{eq:EqM_N2}
\end{eqnarray}
where $\dot{x}=v_x= {p_x}/{m\gamma}$, $\dot{y}=v_y=
{p_y}/{m\gamma}$, and $\dot{f} \equiv df/dt$ for any variable $f$.

An integral of motion $I_0$ can be derived from
Eqs.~(\ref{eq:EqM_N1},\ref{eq:EqM_N2}):
\begin{equation}\label{eq:Energy}
  I_0 = \gamma - \frac{p_x v_{\rm ph}}{mc^2} + \frac{\omega_{p}^2 y^2}{4c^2} +
  \frac{eE_{\parallel} v_{\rm ph}}{\omega_L mc^2}\varphi,
\end{equation}
and the definition of $\gamma=(1+p_x^2/m^2c^2+p_y^2/m^2c^2)^{1/2}$
can be used in combination with Eq.~(\ref{eq:Energy}) to express
$p_y$ in terms of $(\phi,p_x)$, thereby eliminating the transverse dynamics given by Eq.(\ref{eq:EqM_N2}). In the paraxial
approximation ($p_x^2>>p_{y}^2>>m^2c^2$), the transverse
energy~\cite{kostyukov_pop04,malka_prl11} $\varepsilon_{\perp}=
\frac{1}{2 c}|p_x|\dot{y}^2+\frac{1}{4}m\omega_{p}^2  {y}^{ 2}$
can be conveniently expressed as
\begin{equation}\label{eq:transverse_energy}
    \varepsilon_{\perp} = mc^2I_0 -
    \frac{eE_{\parallel} v_{\rm ph}}{\omega_L} \varphi +
    v_{ph}p_x-c|p_x|.
\end{equation}

Equation (\ref{eq:transverse_energy}) is greatly simplified in the case of a luminous laser pulse with $v_{\rm ph} = c$. After taking into account that $\dot{\varphi} \propto ( v_x/c-1) < 0$ and assuming that the electrons are accelerated in the forward direction by the laser, i.e. $p_x>0$, it follows that $\dot{\varepsilon}_{\perp} \propto E_{\parallel}$. This result implies that, regardless of the amplitude of the laser, the {\it accelerating} longitudinal field ($-eE_{\parallel}>0$) depletes the energy of transverse oscillations $\varepsilon_{\perp}$, thereby suppressing the DLA mechanism. Therefore, effective synergy of DLA  and longitudinal field acceleration can be achieved  only  for those electrons with large initial transverse energy~\cite{zhang_prl}, thus constraining the injection.

\begin{figure}[b!]
\includegraphics[height=0.13\textheight,width=.9\columnwidth]{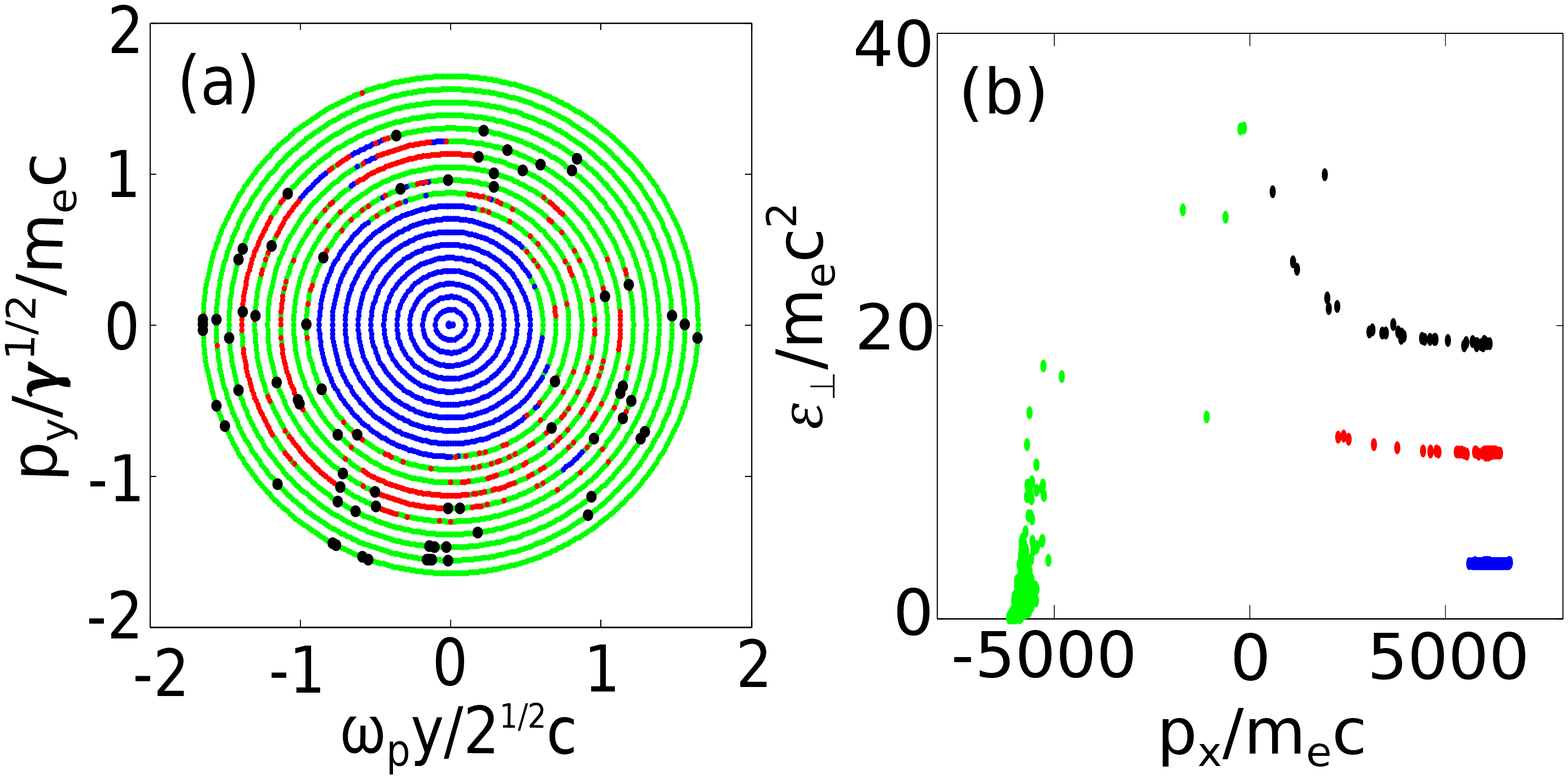}
\caption{\label{fig:gamma_t0}(color online)
 Direct laser acceleration of test electrons in an ion channel by the combination of a laser pulse propagating with the speed of light, and a uniform decelerating electric field. (a) Color-coded energy gain as a function of the initial conditions in the transverse $(y,p_y)$ phase plane. (b) Four distinct groups of accelerated electrons in the $(p_x,\varepsilon_{\perp})$ phase space: backward-accelerated electrons (green, no laser contribution); forward-accelerated electrons by the laser pulse resonant with the first (blue), third (red), and fifth (black) harmonics of the betatron motion. Propagation distance: $x=10^5\lambda_L$, other parameters: in the text.}
\end{figure}

A very different dynamics, which is the subject of this paper,
emerges in the case of the {\it decelerating} longitudinal field
($-eE_{\parallel} < 0$), where two scenarios can be realized: (1)
$\varepsilon_{\perp}$ increases for the forward-accelerated
electrons, and (2) $\varepsilon_{\perp}$ decreases for the electrons accelerated in the backward direction ($p_x<0$) by the
longitudinal field. Because the DLA mechanism is
suppressed in the latter case, we concentrate on the scenario (1), where the  longitudinal  field pumps energy into transverse
betatron oscillations, thereby enabling steady-state acceleration
of the particle by the laser wave against the decelerating force
$-eE_{\parallel}$.

The effect of the decelerating longitudinal electric field
enhancing the energy of the transverse motion is counter-intuitive
because, while the two degrees of freedom are decoupled for
non-relativistic particles, that is no longer the case for
ultra-relativistic electrons. Specifically, by re-writing
Eq.~(\ref{eq:EqM_N2}) without the laser field in a more
conventional form~\cite{cipiccia,cary_prl} as $\ddot{y} + \Gamma
\dot{y} + \omega_{\beta}^2 y = 0$, where
$\omega_{\beta}=\omega_p/\sqrt{2\gamma}$ is the betatron
frequency, and $\Gamma=\dot{\gamma}/\gamma$ is the damping rate
due to relativistic mass increase. Because in the absence of the
laser $\dot{\gamma} \approx -eE_{\parallel}/mc$, we find that $\Gamma<0$ for decelerating electric field, resulting in the growth of betatron oscillations.

To investigate the effect of combining the decelerating
longitudinal field with that of a laser, we have propagated an
ensemble of test electrons from $x=0$ to $x=10^5\lambda_L$ according to Eqs.~(\ref{eq:EqM_N1},\ref{eq:EqM_N2}). The test electrons were initialized with $p_x=25mc$ and uniformly loaded into a transverse phase space corresponding to $0 < \epsilon_{\perp 0} < 1.6mc^2$ as shown in Fig.~\ref{fig:gamma_t0}, where particles are color-coded according to gained energy. The simulation parameters $a_0 \equiv eE_0/m\omega_Lc = 2.12$,
$a_{\parallel} \equiv eE_{\parallel}/m\omega_Lc=0.01$, and
$\omega_p=\omega_L/30$ correspond to the laser intensity
$I_L=10^{19}$W/cm$^2$, the decelerating field of $0.4$GeV/cm, and
the electron plasma density $n=2\times 10^{18}$cm$^{-3}$ for the
laser wavelength $\lambda_L=0.8\mu$m.

The electron phase space undergoes a complicated fragmentation
shown in Figs.~\ref{fig:gamma_t0}(a,b), where four distinct
color-coded groups of electrons can be identified. The
backward-moving group of particles with the final momentum $p_x
\equiv - p_{\rm fin} \sim -6\times 10^3 mc$ group (green)
corresponds to scenario (2), while the other three groups with
$p_x \sim p_{\rm fin}$ that are color-coded according to their
increasing $\epsilon_{\perp}$ as blue, red, and black correspond
to scenario (1). Because the work done by the longitudinal force
on the forward-moving particles is negative and equal to
$\Delta A_W  \approx -c p_{\rm fin}$, by implication the work
done directly by the laser via the DLA mechanism is $\Delta A_{L} = -2\Delta A_W$. Note that even though $E_{\parallel} << E_0$, the longitudinal field has profound stabilizing effect on the direct laser acceleration. Specifically, numerical integration of the Eqs.~(\ref{eq:EqM_N1},\ref{eq:EqM_N2}) with $E_{\parallel}=0$
and the same other conditions yield a much smaller peak electron
momentum $p_x \sim 1.2\times 10^3mc$.

\section{Analytical theory}\label{SecA}

To understand the physics and the limits of the direct laser
acceleration in the presence of the decelerating electric field,
we develop below an analytic theory that assumes (i) linear laser
polarization, (ii) luminal phase velocity $v_{\rm ph}=c$, and (iii) ultra-relativistic forward-accelerated electrons with $p_x \gg mc$. The super-liminal case ($v_{\rm ph}>c$) is briefly analyzed towards the end of this paper, and the case of finite ellipticity of the laser field is addressed in the Appendix~A. For convenience, we replace the longitudinal degrees of freedom $(x,p_x)$ by $(\varphi,\tilde{p}_x \equiv p_x/mc)$, and use the energy-angle variables $I(t) \equiv \varepsilon_{\perp}/mc^2$ and the betatron oscillation phase $\psi$ to express the betatron oscillation as $y= 2(c/\omega_p)\sqrt{I} \cos\psi $, where $\dot{\psi}=\omega _{\beta}$, and we have assumed that in the
ultra-relativistic paraxial limit $\gamma \approx \tilde{p}_x$.
Assuming that the variation of $\tilde{p}_x$ during one betatron
period is small (i.e. $\langle \dot{\psi}\rangle \equiv \langle
\omega_{\beta}\rangle \approx \omega _{\beta}$, where $\langle
\cdot \rangle$  denotes averaging over one betatron period), the
transverse velocity is given by $v_y = v_{*} \sin\psi$, where
$v_{*}/c=\sqrt{2I/\tilde{p}_x}$.

Next, we introduce the time-averaged equation of motion by
averaging the Doppler-shifted laser frequency in the electron's
reference frame defined as $\dot{\varphi}= -\omega_D$, where
$\omega_D = \omega_L(1-v_x/v_{ph}) \approx
\omega_L(v_{*}^2/2c^2)\sin^2{\psi}$ for $v_{\rm ph}=c$. In the
paraxial approximation, the time-averaging of $\langle
\dot{\varphi}\rangle \equiv - \bar{\omega}_{D}$ yields
$\bar{\omega}_{D} = \omega_L I/(2\tilde{p}_x)$. Neglecting the
difference between $\langle \tilde{p}_x \rangle$ and
$\tilde{p}_x$, the time-averaging of Eq.~(\ref{eq:EqM_N1}) yields
$\dot{\tilde{p}}_x =-\omega_L a_{\parallel} - \omega_L a_0 (v_*/c) \eta$. The time-averaged laser-particle interaction strength $\eta = \langle \sin\psi \cos\varphi \rangle$ does not vanish upon averaging only if the resonance condition $\bar{\omega}_D \approx l \omega_{\beta}$ is satisfied for at least one odd betatron harmonic, in which case $\eta \approx \alpha_l \sin\theta_l$, where $\theta_l = \langle \varphi \rangle + l\psi$ is the phase detuning between the laser field and the odd-integer $l$'th harmonic of the betatron motion (see Appendix~B and \cite{Davoine2014}). The largest non-vanishing coefficients are $\alpha_1 \approx 0.35$, $\alpha_3=0.16$, and $\alpha_5=0.11$ .

To further simplify the calculation, we assume the resonance with
the $l$'th betatron harmonic and express the averaged equations of motion as
\begin{eqnarray}
 \dot{\tilde{p}}_x &=&-\omega_L a_{\parallel} \left( 1 + \frac{2v_{*}}{v_{\rm cr}^{(l)}} \sin\theta_l \right), \ \ \dot{\theta}_l = l\omega_{\beta} - \bar{\omega}_D \label{eq:EqM_F1}\\ \dot{I} &=& I \frac{a_{\parallel}\omega_L}{2\tilde{p}_x},
  \ \ \dot{\psi} = \omega_{\beta}, \label{eq:EqM_F2}
\end{eqnarray}
where Eq.~(\ref{eq:transverse_energy}) was used to express
$\dot{I}$, and $v_{\rm cr}^{(l)} = 2ca_{\parallel}/\alpha_la_0$ is the critical betatron velocity such that, as shown below, the
direct laser acceleration stops for $v_{*} < v_{\rm cr}^{(l)}$. For a resonant electron with the $(\tilde{p}_x,I)=(p_r,I_r)$
momentum/energy satisfying
$\bar{\omega}_D(\tilde{p}_r,I_r) = l \omega_{\beta}(\tilde{p}_r)$
the following relationship must hold for all times: $I_r/\sqrt{2p_r} = l\omega_p/\omega_L$. Therefore, after integrating
Eq.~(\ref{eq:EqM_F2}) we obtain the following solution:
\begin{equation}\label{eq:EqR_1}
  \tilde{p}_r(t) = \tilde{p}_{0}+ a_{\parallel}\omega_L t, \ \
  \frac{v_r}{c} = \left( \frac{l \omega_p}{\omega_L} \right)^{1/2}
  \left( \frac{2}{p_r(t)}\right)^{1/4},
\end{equation}
where the slow decrease of the resonant electron's transverse
velocity $v_r \equiv v_{*}(I_r,p_r)$ is compensated by the slow
drift of its phase according to $\sin{\theta_r^{(l)}} = -v_{\rm cr}^{(l)}/v_r(t)$ to ensure the constant-gradient acceleration represented by Eq.~(\ref{eq:EqR_1}).

It immediately follows from Eq.~(\ref{eq:EqR_1}) that a resonant
particle gains energy via the DLA mechanism at the rate
that is twice the rate of energy loss to the decelerating
longitudinal field regardless of the order of the betatron
resonance, i.e. $\Delta A^{(L)} = -2\Delta A^{(W)}$ for any value
of $l\geq 1$. However, the resulting transverse energy gains
$\varepsilon_{\perp} \propto l$. This scaling explains the
physical reason for the fragmentation of the electron phase space
observed in Fig.~\ref{fig:gamma_t0}(b): the transverse energy
"plateaus" of accelerated electrons with different values of
$\varepsilon_{\perp}$ correspond to different resonant betatron
harmonics. Moreover, it follows from $|\sin{\theta_r^{(l)}}| \leq
1$ that the maximum energy gain corresponding to
$\sin{\theta_r^{(l)}} = -1$ depends on the harmonic number:
\begin{equation}\label{eq:EqL_1}
    \tilde{p}_{\max}^{(l)} = \frac{l^2 \alpha_l^4}{2}
    \left( \frac{\omega_p}{\omega_L} \right)^2
    \left( \frac{E_0}{E_{\parallel}} \right)^4.
\end{equation}
Because the quantity $l^2\alpha_l^4$ decreases with $l$, the
following hierarchy of energy gains is established between the
three lowest harmonics: $\tilde{p}_{\max}^{(1)} >
\tilde{p}_{\max}^{(3)} > \tilde{p}_{\max}^{(5)}$.

\begin{figure}[t!]
\includegraphics[height=0.26\textheight,width=.9\columnwidth]{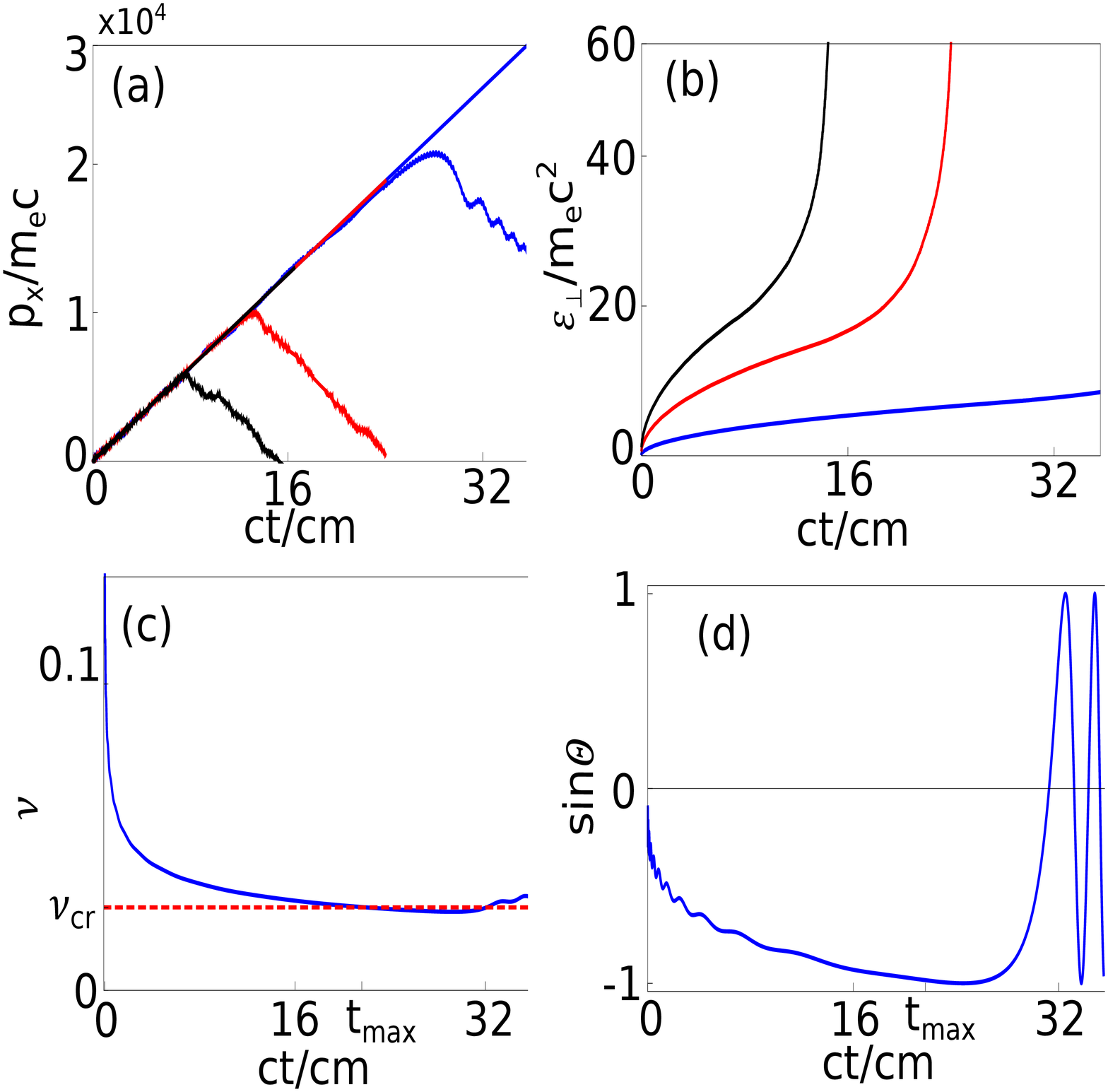} \caption{\label{fig:gamma_t1}(color online)
 The dynamics of the representative accelerated electrons from the three groups in Fig.~\ref{fig:gamma_t0} corresponding to first (blue), third (red), and fifth (black) sub-harmonics of the Doppler-shifted laser frequency $\omega_D$ resonantly interacting with electrons' betatron motion in an ion channel: (a) the normalized longitudinal momentum $p_x/mc$, and (b) transverse energy  $\varepsilon_{\perp}$. (c) Betatron velocity amplitude $v_{*}$ and (d) phase mismatch $\theta$ of the test particle from the first-harmonic resonance group. Constant-gradient acceleration stops when $\sin\theta=-1$ and $v_{*}=v_{\rm cr}$ ( dashed line in (c)).}
\end{figure}

To verify these analytic predictions, three test particles labeled
as blue ($l=1$), red ($l=3$), and black ($l=5$) are selected from
the phase space shown in Fig.~\ref{fig:gamma_t0}(a) and propagated
up to $x=4\times 10^5\lambda_L$ (i.e. over a $4$ times longer
distance than in Fig.~\ref{fig:gamma_t0}). The electrons from all
three groups are clearly accelerated at the same constant rate as
shown in Fig.~\ref{fig:gamma_t1}(a). The main distinction between
the three groups is their gained transverse energy:
$d\varepsilon_{\perp}/dt$ increases with the harmonic number as
shown in Fig.~\ref{fig:gamma_t1}(b). The amplitude $v_{*}$ of the
betatron velocity, which is extracted from $v_y(t)$ by
time-averaging, is shown in Fig.~\ref{fig:gamma_t1}(c). As
predicted by Eq.(\ref{eq:EqR_1}), $v_{*}$ decreases with time, and
the accelerating gradient is kept up at its constant value by the
increase in $|\sin{\theta}|$ as shown in
Fig.~\ref{fig:gamma_t1}(d). However, when $v_{*} = v_{\rm
cr}^{(l)}$ and $\sin{\theta}=-1$ conditions are reached,
constant-gradient acceleration stops, and the electron's energy
starts rapidly decreasing after the critical time $t=t_{\rm max}$
as shown in Fig.~\ref{fig:gamma_t1}(a).

To understand the physics of the abrupt transition at $t=t_{\max}$
from constant-gradient acceleration to rapid deceleration, we
linearize Eqs.~(\ref{eq:EqM_F1}) around the resonant particle's
momentum according to $\tilde{p}_{x}=\tilde{p}_{r} + \delta
\tilde{p}$ (here we assume $l=1$ and drop the harmonic label). A
pendulum-like equation is obtained (see Appendix~B for the details of the linearization
procedure):
\begin{equation}\label{eq:EqM_F210011}
    \frac{d}{d t}M_I\frac{d\theta}{d t} = -T \left(
    \frac{v_{cr}}{v_r(t)} + \sin{\theta} \right),
\end{equation}
where $M_I=I^3{\omega_L^2}/{\omega_p^4}$ and $T ={2\alpha_1 a_0 \omega_p}/({I^{1/2}\omega_L})$. Note that $v_r(t) >
v_{\rm cr}$ for early $t<t_{\rm max}$ times, so that
Eq.~(\ref{eq:EqM_F210011}) has a slowly evolving stable equilibrium point $\theta(t) = -\sin^{-1}(v_{cr}/v_r)$. The equilibrium point disappears at $t=t_{\max}$ when the phase reaches $\theta=-\pi/2$ which separates the oscillation and rotation regions of the pendulum's phase space.

Therefore, for $t > t_{\rm max}$ the pendulum starts rotating with an increasing speed under the action of a constant torque. In the context of electron acceleration, this implies that, while the amplitude of the Lorentz force is still large because the speed of the betatron oscillations remains close to $v_{\rm cr}$ as shown in Fig.~\ref{fig:gamma_t1}(c), the fast oscillations of
the mismatch phase $\theta$ result in the vanishing of the
time-averaged direct laser acceleration. This causes electron's overall deceleration by the longitudinal force $-eE_{\parallel}$ as shown in Fig.~\ref{fig:gamma_t1}(a) for $t > t_{\rm max}$.

\section{Simulations}\label{subsec::5a}
The above simplified model assumes a constant in time decelerating longitudinal field. In reality such fields cannot be maintained over long distances, and a time-changing longitudinal fields must be employed. One of the most promising approaches to producing such fields is a laser-wakefield accelerator (LWFA) concept, where significant experimental progress has been recently achieved~\cite{faure_nature04,geddes_nature04,mangles_nature04,
leemans_1gev,wang_2gev,leemans_4gev}. Below we demonstrate that constant-gradient direct laser acceleration can be achieved inside a plasma bubble, where externally injected electrons directly interact with the bubble-forming laser pulse and with the decelerating longitudinal field in the front portion of the bubble.

\begin{figure}[t]
\centering
\includegraphics[height=0.26\textheight,width=.9\columnwidth]{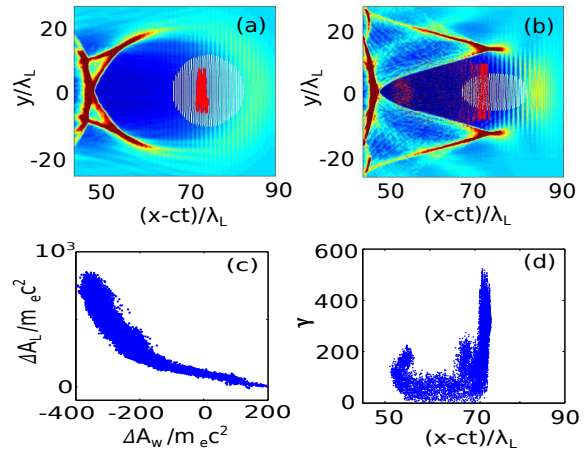}
\caption{PIC simulations of direct laser acceleration of electrons injected
into the decelerating phase of a plasma bubble in a LWFA. Plasma density
(color-coded), externally injected electron beam (red dots), and laser
intensity contour (white) at (a) $ct = 120\lambda_L$  and (b) $ct = 700\lambda_L$.
(c) Electron energy gains from the wake ($\Delta A_W$) and from the laser
($\Delta A_L$) fields. (d) ($x-ct$,$\gamma$) phase space of injected  electrons
at $t=700\lambda_L$. Plasma density $n = 4.3 \times 10^{18} cm^{-3}$, beam's
density $n_{b} = 4.3 \times 10^{15} cm^{-3}$ and initial momentum $p_{x0}=25 mc$.
Laser parameters: wavelength $\lambda_L = 0.8 \mu m$, intensity $I = 7.7 \times 10^{19} W/cm2$,
pulse duration $\tau = 35 fs$, spot size $w_0 = 12 \mu m$.}\label{fig:simulation}
\end{figure}

Two-dimensional in space and three-dimensional in electron
velocity simulations using a VLPL particle-in-cell (PIC)
code~\cite{pukhov_vlpl} were carried out for the laser parameters
listed in the caption of Fig.~\ref{fig:simulation}. A short
electron bunch with the duration $\tau_b = 5\lambda_L/c$ injected
with the initial momentum $p_x=25mc$ co-propagates with the laser
pump pulse as shown in Fig.~\ref{fig:simulation}(a). The front of
the bunch initially coincides with the peak of laser pulse so that
the beam is positioned in the region where the combined effect of
the plasma wakefield and of the laser's ponderomotive force is to
decelerate the injected electrons. Although the characteristics of
the laser pulse and its bubble evolve (see
Figs.~\ref{fig:simulation}(a) and \ref{fig:simulation}~(b) for
comparison),  the  majority  of the injected electrons experience
a decelerating wakefield throughout the simulation's duration of
$t=700\lambda/c$.

The energies gained from the wake, $\Delta A_W = - \int e E_x v_x
dt$, and from the DLA mechanism, $\Delta A_L = - \int e
E_y \cdot v_y dt$, were estimated~\cite{gahn_prl,zhang_prl} by
extracting $E_{x,y}$ from the PIC simulations and integrating it
for each injected electron over a distance of $ct \approx 0.56$mm. The results shown in Fig.~\ref{fig:simulation}(c) clearly indicate that those electrons that have lost energy to the wake ($\Delta A_W < 0$) have gained energy from the laser ($\Delta A_L > 0$), and that $\Delta A_L \approx -2 \Delta A_W$ for the highest energy electrons as predicted by the analytic theory. For example, those
electrons that have gained $\Delta A_L \approx 750mc^2$ from the
DLA mechanism have lost $\Delta A_W \approx 350mc^2$ to
the decelerating wakefield, thereby gaining over $200$MeV in
energy.

The high-energy electrons roughly correspond to the blue- and
red-coded test particles in Fig.~\ref{fig:gamma_t0}(b), with some
of the electrons undergoing a transition from $l=1$ to $l=3$
resonance during the propagation. Their physical location inside
the bubble at $x-ct \approx 70 \lambda_L$ can be identified in
Fig.~\ref{fig:simulation}(b) by their large betatron amplitude,
and in Fig.~\ref{fig:simulation}(d) by their large total energy.
On the other hand, the low-energy electrons roughly corresponding
to the green-coded test particles in Fig.~\ref{fig:gamma_t0}(b) do not directly interact with the laser pulse. Instead, driven by the decelerating wakefield, they initially slip to the back of the
bubble at $x-ct \approx 50 \lambda_L$ where the wakefield changes
sign, and subsequently gain a small amount of energy directly from the wake. Since no energy gain from the DLA mechanism
takes place for these electrons, their betatron amplitude is strongly reduced as can be observed in Fig.~\ref{fig:simulation}(b).

\section{Discussions}

Note that the phase velocity $v_{\rm ph}$ of the laser field
inside the plasma bubble is slightly higher than the speed of
light $c$ despite the low density of the ambient plasma ($\omega_p
\ll \omega_L$) and still lower plasma density inside the plasma
bubble. Therefore, at least some deviations from the idealized
acceleration scaling given by Eq.~(\ref{eq:EqR_1}) due to finite
$\delta v_{\rm ph} = v_{\rm ph} - c$ is expected. The
super-luminal effects are negligible as long as $\delta v_{\rm ph}
\ll \left( c - \langle v_x \rangle \right)$, and
Eq.~(\ref{eq:EqL_1}) holds if $(E_{\parallel}/E_0)^2 \gg
\alpha_l^2 \delta v_{\rm ph}/c$ for the $l$'th harmonic
acceleration. In the opposite limit, maintaining the betatron
resonance imposes a slightly modified relationship between
transverse energy and longitudinal momentum: $I= l \sqrt{{2\tilde
p_{x}}}\omega_p/\omega_L - 3\delta v_{\rm ph} \tilde{p}_x/c$. A
modified condition for the maximum electron energy can be obtained
from $dI/d\tilde{p}_x=0$: $\tilde{p}_{x,\max}^{(l)}\sim
(l\omega_p/\omega_L)^2/18(\delta v_{\rm ph}/c)^2$.

In conclusion, we have proposed a novel approach to
constant-gradient far-field particle acceleration: an direct laser acceleration combined with longitudinal deceleration. The
combination of large transverse momentum and large total energy
makes such accelerators promising for developing compact radiation
sources. Technological advances in synchronizing electron bunches
and laser pulses ensure that the suggested scheme of injecting
electrons into a decelerating phase of a plasma bubble will be
experimentally realized. Future work will explore the possibility
of extending the range of accelerated energies by adiabatically
varying plasma parameters.

 \section{Acknowledggements}
This work was supported by DOE grants DE-SC0007889 and
DE-SC0010622, and by AFOSR grant FA9550-14-1-0045. The authors thank the
Texas Advanced Computing Center (TACC) at The University of Texas at Austin for providing HPC resources.

 \appendix

\section{The resonant acceleration of electrons by elliptically polarized  laser wave.}

\begin{figure}[b!]
 \centering
 \includegraphics[height=0.300\textheight,width=0.9\columnwidth]{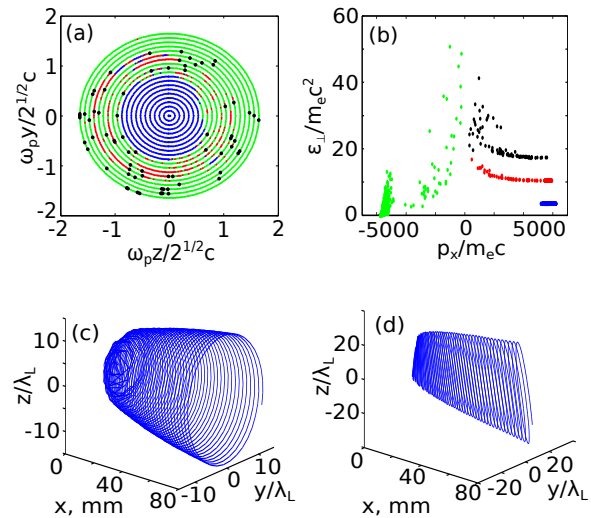}
 \caption{Direct laser acceleration of test electrons in an ion channel by the combination of an elliptically polarized laser pulse ($E_1/E_2=3$) propagating with the speed of light, and a uniform decelerating field.  (a) Color-coded gain of the longitudinal momentum $p_x$ as a function of the initial conditions in the transverse $(y, z)$  plane. The initial  momentum $p_x=25m_ec$, and the initial transverse momenta $p_y=p_z=0$.  (b) Four distinct groups of accelerated electrons in the $(p_x,\varepsilon_{\perp})$ phase space: backward-accelerated electrons (green, no laser contribution); forward-accelerated electrons by the laser pulse resonant with the first (blue), third (red), and fifth (black) harmonics of the betatron motion. Propagation distance:  $x=10^5\lambda_L$, other parameters are the same as in the case presented in the Fig.~1 in the main manuscript. Trajectories of the representative (c) blue-colored ($l=1$) and (d) red-colored particle ($l=3$) electrons. }\label{fig:3d_trajectory}
\end{figure}

In the main manuscript we assumed that the laser pulse interacting with betatron-oscillating electrons is linearly polarized. This assumption was made in order to simplify the calculation. Here we show that any elliptically polarized laser provides constant-gradient direct laser acceleration whenever an additional {\it decelerating} longitudinal electric field is introduced.
The elliptically polarized laser wave is assumed to be planar $E_y^{(L)} = E_1 \cos{\phi}$, $E_z^{(L)} = E_2 \sin{\phi}$ and $B_z^{(L)} =E_y^{(L)}$ and $B_y^{(L)} =- E_z^{(L)}$, where $\phi=\omega_L (x/c - t)$. Relativistic equations of motion are given by
\begin{eqnarray}
&&\hspace{-10mm}\frac{d}{dt} \left( \gamma \frac{dx}{dt} \right) = -\frac{e}{m} \left( E_{\parallel} + E_1
\frac{v_y \cos{\varphi}}{c}+E_2\frac{v_z \sin{\varphi}}{c} \right), \label{eq:AEqM_N1} \\
&&\hspace{-10mm}\frac{d}{dt} \left(\gamma \frac{dy}{dt} \right) = -\frac{\omega_p^2 y}{2}
+ \frac{eE_1}{m} \left( \frac{v_x}{c} - 1 \right) \cos{\varphi}, \label{eq:AEqM_N2}\\
&&\hspace{-10mm}\frac{d}{dt} \left(\gamma \frac{dz}{dt} \right) = -\frac{\omega_p^2 z}{2}
+ \frac{eE_2}{m} \left( \frac{v_x}{c} - 1 \right) \sin{\varphi}. \label{eq:AEqM_N3}
\end{eqnarray}

We illustrate the interaction of test particles with the elliptically polarized wave considering the case when the ratio between two orthogonal components of the electric field  equal to $3$ ($E_1/E_2=3$). The wave intensity and all other parameters are  the same as in the case shown in Fig.~1 in the main manuscript.

The integration of the Eqs.~(\ref{eq:EqM_N1}) - (\ref{eq:EqM_N3}) shows that  initially homogeneously seeded  phase space  undergoes strong fragmentation (see Figs.~\ref{fig:3d_trajectory}(a,b) for electron phase spaces color-coded in accordance with their final change of the longitudinal momentum) in the same manner as in the case of the linear polarized laser wave shown in Fig.1 of the main manuscript. With the exception of green-colored particles that gain negative momentum $p_x < 0$ from the decelerating electric field, all other electrons gain transverse energy $\varepsilon_{\perp}$ and positive longitudinal momentum $p_x>0$ via the DLA mechanism. 

In contrast with the linear polarization case, where the electrons were moving mostly in the ($x-y$) plane, they now move along more complex three-dimensional trajectories. Specifically, the electrons gyrate in the $(y-z)$ plane while propagating in the longitudinal ($x-$) direction. Our numerical simulations indicate that the exact nature of the electron's gyration depends on the nature of the harmonic order $l$ of the betatron resonance. Here, as in the main manuscript, we assume that $\Delta\bar{\omega}_D = l \omega_{\beta}$, where $\Delta\bar{\omega}_D$ is the time-averaged Doppler shifted laser frequency, and $\omega_{\beta}$ is the betatron frequency. 

For example, we find that an electron's trajectory is almost circular in the $(y, z)$ plane if the laser wave is resonant with the  first ($l=1$) harmonic of the betatron motion. An example of such trajectory is shown in Fig.~\ref{fig:3d_trajectory}(c). It is noteworthy that the circular nature of the electron's gyration is preserved despite the strong ellipticity of the laser pulse.
On the other hand, the trajectory can become highly elliptical when the laser wave is resonant with the  third ($l=3$) harmonic of the betatron motion. An example of such a trajectory, which is strongly elongated in the $z-$direction with the axis ratio $\sim 9$, is shown in Fig.~\ref{fig:3d_trajectory}(d). Note that this elongation takes place despite the fact that the laser's polarization ellipse is elongated in the $y-$direction. The ellipsis are squeezed even more for the black-colored ($l=5$) particles.

\section{Averaging procedure for Lorentz force.}
Let us consider the luminal case when $\chi=0$ and assume that the longitudinal momentum changes slowly with time  and calculate. Near the resonance $\langle{\omega}_D\rangle=l\omega_{\beta}$, the particle oscillations' phase and the wave phase satisfy equations:
$\dot{\psi} =\omega_{\beta}$ and $\dot{\varphi} =-2l\omega_{\beta}{\sin^2\psi}$ so that
\begin{eqnarray}
 l\psi+\varphi= \langle\theta\rangle +\frac{l}{2}\sin {2 \psi},\label{eq:SM1}
\end{eqnarray}
where $\langle\theta\rangle=l\langle\psi\rangle+\langle\phi\rangle$, and averaging is performed over betatron phase $\psi$. In the Lorentz force we average
\begin{eqnarray}
 \frac{1}{{2 \pi}}\int_{0}^{2 \pi}\sin\psi\cos\varphi d\psi=\nonumber\\
\frac{1}{{2 \pi}}\int_{0}^{2 \pi}\sin\psi\cos\Big( \langle\theta\rangle-l\psi +\frac{l}{2}\sin {2 \psi}\Big)d\psi=\nonumber\\
\alpha_l\sin\langle\theta\rangle,\label{eq:SM2}
\end{eqnarray}
where
\begin{eqnarray}
\alpha_l= \frac{1}{{2 \pi}}\int_{0}^{2 \pi}\sin\psi\sin\Big(l\psi -\frac{l}{2}\sin {2 \psi}\Big)d\psi=\nonumber\\
\frac{1}{2}[1-(-1)^l]\Big[J_{l_1}\Big(\frac{l}{2}\Big)-J_{l_2}\Big(\frac{l}{2}\Big)\Big].\label{eq:SM3}
\end{eqnarray}
where $l_{1,2}=(l \mp 1)/2$. Thus, $\alpha_1=0.348101$, $\alpha_2=0$, $\alpha_3=0.162924$,  $\alpha_4=0$, and $\alpha_5=0.114729$.

\section{Similarity with the motion of pendulum.}
Averaged equations of motion do not depend on time explicitly
\begin{eqnarray}
 \dot{\tilde{p}}_x =-\omega_L a_{\parallel}-\alpha_l\omega_L a_0 ({2I}/{\tilde{p}_x })^{1/2}\sin\theta_l  ,\\
{\dot{\theta}}=-\omega_L (I/2\tilde{p}_x)+l\omega_{p}/(2\tilde{p}_x )^{1/2}, \label{eq:EqM_F2}\\
 \dot{{I}}=a_{\parallel}\omega_L (I/2\tilde{p}_x).
\label{eq:EqM_F3}
\end{eqnarray}
It is convenient to use the transverse energy $I$ as an independent variable. Dividing Eqs.~(\ref{eq:EqM_F1}) and (\ref{eq:EqM_F2}) by Eq.~(\ref{eq:EqM_F3}), we obtain for $l=1$

\begin{eqnarray}
 \frac{a_{\parallel}I}{2\tilde{p}_x}\frac{d{\tilde{p}}_x}{dI} =- a_{\parallel}-\alpha_1 a_0 \Big(\frac{2I}{\tilde{p}_x }\Big)^{1/2}\sin\theta ,\label{eq:EqM_F11}\\
-a_{\parallel}\frac{d{\theta}}{dI}=1-\frac{\omega_{p}}{\omega_L}\frac{\sqrt{2\tilde{p}_x }}{I}. \label{eq:EqM_F21}
 \end{eqnarray}
During steady resonant acceleration $a_{\parallel}{d{\theta}}/{dI}\ll 1$ and in the zeroth approximation we obtain from Eqs.~(\ref{eq:EqM_F21}) and (\ref{eq:EqM_F11}):
 \begin{eqnarray}
 \tilde{p}_{r} =\frac{1}{2}\frac{{\omega_L}^2}{\omega_p^2}I^2,\label{eq:EqM_p0}\\
\sin\theta_r=-\frac{a_{\parallel}I^{1/2}}{\alpha_1 a_0(\omega_p/\omega_L)}. \label{eq:EqM_theta0}
 \end{eqnarray}
To analyze the solution of Eq.~(\ref{eq:EqM_F1}) - (\ref{eq:EqM_F3})  beyond the point where formally $|\sin\theta_r|$ becomes greater than $1$, we consider the next approximation by presenting $\tilde{p}_{x}=\tilde{p}_r+\delta \tilde{p}_{x}$.
Substitution of this expansion into  Eqs.~(\ref{eq:EqM_F11}) and (\ref{eq:EqM_F21}) yields
\begin{eqnarray}
\frac{a_{\parallel}I}{2\tilde{p}_r}\frac{d{\delta\tilde{p}}_x}{dI}=\nonumber\\
\big[- 2a_{\parallel}-2\alpha_1 a_0 (\omega_p/\omega_L)I^{-1/2}\sin\theta\big]\bigg(1-\frac{1}{2}\frac{\delta \tilde{p}_{x}}{\tilde{p}_r}\bigg),\label{eq:EqM_F21001} \\
-a_{\parallel}\frac{d{\theta}}{dI}=-\frac{1}{2}\frac{\delta\tilde{p}_x }{\tilde{p}_r}.\,\, \label{eq:EqM_F21000}
 \end{eqnarray}
Since expression into square brackets in the r.h.s. of Eq.~(\ref{eq:EqM_F21001}) is small near the zeroth approximation, we can replace the factor $(1-{\delta \tilde{p}_{x}}/{\tilde{p}_r})$ by $1$. Then combining Eqs.~(\ref{eq:EqM_F21001}) and (\ref{eq:EqM_F21000}), we obtain
\begin{eqnarray}
a_{\parallel}^2\frac{1}{I}\frac{d}{dI}I^3\Big(\frac{1}{I}\frac{d\theta}{dI}\Big)=\nonumber\\
- 2a_{\parallel}-2\alpha_1 a_0 (\omega_p/\omega_L)I^{-1/2}\sin\theta,\label{eq:EqM_F210011}
 \end{eqnarray}
Introducing new 'time'  $\tilde t$ by formula: $IdI=(\omega_p/\omega_L)^2d\tilde{p}_r=a_{\parallel}(\omega_p/\omega_L)^2\omega_L d\tilde t$, we can transform last equation to the form:
\begin{eqnarray}
\frac{\omega_L^2}{\omega_p^4}\frac{d}{d\tilde t}I^3\frac{d\theta}{d\tilde t}=
- 2a_{\parallel}-2\alpha_1 a_0 (\omega_p/\omega_L)I^{-1/2}\sin\theta.\label{eq:EqM_F210011}
 \end{eqnarray}
Finally let us not that we can return to regular time $t$ because difference between $\tilde t$ and $t$ is small.


  \nocite{*}



\end{document}